\documentclass{article}

\usepackage{arxiv}

\usepackage[utf8]{inputenc} 
\usepackage[T1]{fontenc}    
\usepackage{hyperref}       
\usepackage{url}            
\usepackage{booktabs}       
\usepackage{amsfonts}       
\usepackage{nicefrac}       
\usepackage{microtype}      

\usepackage{amssymb}

\usepackage{lineno}

\usepackage{latexsym}

\usepackage[english]{babel}
\usepackage{enumerate}

\usepackage{array}

\usepackage{graphicx}
\usepackage{caption}
\usepackage{subcaption}

\usepackage{float}
\usepackage{amsthm}

\usepackage{amssymb}

\usepackage{dsfont}
\usepackage{mathtools}
\usepackage{tikz-cd}

\usepackage{graphicx}

\usepackage{enumitem}

\makeatletter
\newcommand\mathcircled[1]{%
  \mathpalette\@mathcircled{#1}%
}
\newcommand\@mathcircled[2]{%
  \tikz[baseline=(math.base)] \node[draw,circle,inner sep=1pt] (math) {$\m@th#1#2$};%
}
\makeatother

\usepackage{hyperref}
    \hypersetup{
        colorlinks   = true,
        citecolor    = blue,
        linkcolor=blue  
    }

\newtheorem{theo}{\textbf{Theorem}}[section]

\newtheorem{prop}[theo]{\textbf{Proposition}}

\title{Coevolution of the reckless prey and the patient predator}

\author{
  Cecilia Berardo\thanks{Corresponding author.} \\
  Department of Mathematics and Statistics\\
  FI-00014 University of Helsinki, Finland\\
  \texttt{cecilia.berardo@helsinki.fi} \\
   ORCID: 0000-0002-1729-3765
   \And
 Stefan Geritz \\
 Department of Mathematics and Statistics\\
 FI-00014 University of Helsinki, Finland\\
 \texttt{stefan.geritz@helsinki.fi} \\
 ORCID: 0000-0002-7865-3541}

\begin{document}
\maketitle

\begin{abstract}
The war of attrition in game theory is a model of a stand-off situation between two opponents where the winner is determined by its persistence. We model a stand-off between a predator and a prey when the prey is hiding and the predator is waiting for the prey to come out from its refuge, or when the two are locked in a situation of mutual threat of injury or even death. The stand-off is resolved when the predator gives up or when the prey tries to escape. Instead of using the asymmetric war of attrition, we embed the stand-off as an integral part of the predator-prey model of Rosenzweig and MacArthur derived from first principles. We apply this model to study the coevolution of the giving-up rates of the prey and the predator, using the adaptive dynamics approach. We find that the long term evolutionary process leads to three qualitatively different scenarios: the predator gives up immediately, while the prey never gives up; the predator never gives up, while the prey adopts any giving-up rate greater than or equal to a given positive threshold value; the predator goes extinct. We observe that some results are the same as for the asymmetric war of attrition,
but others are quite different.
\end{abstract}



\section{Introduction}
The war of attrition in game theory is a model of a stand-off situation between two opponents where the winner is determined by its persistence. In the symmetric version of the game, where the costs and benefits for two equally matched opponents are the same, the evolutionarily stable strategy (ESS) is stochastic and given by a negative exponential probability distribution for the length of time till giving-up if the cost of waiting is a linear function of time (\cite{smith1973logic}, \cite{bishop1978generalized}, \cite{smith1982evolution}). The exponential distribution is equivalent to both players adopting the same constant giving-up rate and the average pay-off for each player turns out to be zero.\\ 

In the asymmetric version of the game, where the opponents assume different unambiguous roles like "owner" and "intruder" or "prey" and "predator", there is no ESS under complete information, but a Nash equilibrium where one player gives up immediately while the other player can choose any giving-up time above a certain threshold value (\cite{selten1980note}). This neutrality of strategy choice for the second player can be resolved if the game is even slightly perturbed, e.g., by introducing the possibility of players making errors in the role identification (\cite{hammerstein1982asymmetric}, \cite{kim1993evolutionary}). In some cases, however, errors are unlikely, such as with a stand-off between a predator and its prey.\\

In this paper, we study a stand-off between a predator and its prey, e.g., when the prey is hiding and the predator is waiting for the prey to come out, or more dramatically, when they are locked in a situation of mutual threat of injury or even death by the predator's teeth and claws or the horns and hooves of the prey. The stand-off is resolved when the predator gives up or when the prey tries to escape. More specifically, the predator may give up the prey, which could then escape and survive the attack, or, the prey may have to eventually give up and, as a consequence, be killed by the predator with a certain probability. \\

This behaviour is widely present in nature, as different anti-predator strategies have been observed in prey species upon encounter with a predator in order to maximise survival or, on the other side, some predator species use predatory strategies to lure the prey (\cite{krebs2009behavioural}). Examples are the use of deterrent signals (\cite{fitzgibbon1988stotting}, \cite{favreau2010interactions}), deimatic behaviour (\cite{stevens2005role}, \cite{vallin2005prey},  \cite{ratcliffe2005adaptive}, \cite{langridge2007selective}), playing dead (\cite{alboni2008origin}), physical and chemical features (\cite{vincent1986mechanical}, \cite{maan2012poison}). \\

Studies on the \emph{ecology of fear} include the works by \cite{brown1999vigilance} and \cite{brown2007foraging}, investigating the adaptive behaviour of foragers in order to minimise the cost of predation by selecting the time spent in a certain habitat and the level of vigilance; the experimental works by  \cite{katz2010playing} and \cite{katz2013optimal}, which focus on the cost and benefits of the waiting game in the \emph{little egret-goldfish} dynamics; finally, we cite the articles by \cite{krivan1997dynamic} and \cite{kvrivan2007lotka}, using the Lotka-Volterra system to model the adaptive behaviour of a predator and its prey when the two maximise their fitness by adapting to the other player's strategy or by habitat selection.\\

Instead of using the asymmetric war of attrition, we embed the stand-off as an integral part of the predator-prey model of \cite{rosenzweig1963graphical} derived from first principles. We use the mechanistic method described by \cite{Berardo:2020aa} to derive a predator functional response similar to a Holling type II, where the effective attack rate and effective handling time are interpretable functions of the underlying individual dynamics rates. In this way, the event rates describing individual birth and death, prey capture and prey handling, the formation and break-up of a predator-prey pair in a mutual stand-off, they all become explicit parameters of the population model. The costs and benefits of victory or defeat during a stand-off remain implicit as part of the population dynamics, but eventually they come down to births and deaths gained or lost.\\

In this context, we study the coevolution of the giving-up rates of the predator and the prey using the adaptive dynamics approach (\cite{metz1995adaptive}, \cite{geritz1997dynamics}, \cite{geritz1998evolutionarily}, \cite{geritz1999evolutionary}). As a consequence, our emphasis is somewhat different than in evolutionary game theory. In particular, we focus on the evolutionary dynamics and local stability assuming small mutation steps.\\

As we consider the predator and the prey locked together in the stand-off, our study differs also from the previous works by \cite{geritz2014deangelis}, \cite{lehtinen2019cyclic} and \cite{lehtinen2019coevolution} on evolution of prey timidity as measured by the rates of individual prey entering and leaving a refuge or aggressive defensive posture.\\

The paper is organised as follows. In Section \ref{sec2}, we derive the population equations for the ecological dynamics and the corresponding equilibria. In Section \ref{sec3}, we introduce the adaptive dynamics framework  in the current context and, in Section \ref{sec4}, we study the coevolution of the predator and prey giving up rates. In particular, first we discuss the possible evolutionary phase planes in Section \ref{sec41} and later, in Section \ref{sec42}, we use the canonical equation of adaptive dynamics to understand the direction of evolution.

\section{Ecological dynamics}\label{sec2}
We model the scenario where a predator and a prey species interact in the following way. A searching predator ($y^S$) finds and attacks a foraging prey ($x^F$) at a rate $p$, and with probability $\nu$ the prey is captured and killed, while the predator enters the handling state ($y^H$) which includes eating, digesting, resting and giving birth. With the complementary probability $1-\nu$ the two individuals enter a stand-off state ($P$) where the prey may hide in a refuge or show a high level of alertness or aggressiveness. At the same time the predator does not give up the prey but waits a favourable time for attacking. The stand-off is resolved with rate $q$ when the predator gives up, or with rate $s$ when the prey tries to escape. We name $s$ and $q$ the giving-up rates of, respectively, the prey and the predator. With probability $1-\theta$ the prey successfully escapes and with the complementary probability $\theta$ it is captured and killed after all. \\

The above narration can be summarised with the following fast processes (see also Table \ref{tab:param} for the complete list of model variables and parameters):

\begin{flalign}\nonumber
& \mathcircled{y^S} +\;\mathcircled{x^F}   \xrightarrow{\nu p} \mathcircled{y^H}  \quad   \textit{attack and prey capture},  \\ \nonumber
& \mathcircled{y^S} +\;\mathcircled{x^F}   \xrightarrow{(1-\nu) p} \mathcircled{P}  \quad   \textit{failed attack leading to a stand-off},\\ \nonumber
& \mathcircled{P}   \xrightarrow{q} \mathcircled{y^S}+ \mathcircled{x^F}  \quad  \textit{predator ends stand-off, prey goes free},  \\ \nonumber
& \mathcircled{P}  \xrightarrow{\theta s}  \mathcircled{y^H}  \quad  \textit{prey ends stand-off and is captured}, \\ \nonumber
& \mathcircled{P}  \xrightarrow{(1-\theta ) s}  \mathcircled{y^S}+ \mathcircled{x^F}   \quad  \textit{prey ends stand-off and escapes},\\\nonumber 
& \mathcircled{y^H}  \xrightarrow{ \frac{1}{h} } \mathcircled{y^S}  \quad  \textit{handling predator resumes searching}.
\end{flalign}

Consider $n$ predator types and $m$ prey types. The predators type $j$ with density $y_j$ differ in their giving-up rates $q_j$ and in the same way the prey type $i$ with density $x_i$ and different giving-up rates $s_i$. We define with $P_{ij}$ the density of pairs with predator type $j$ and prey type $i$. From the individual processes we derive the differential equations for the fast time population dynamics as
\begin{eqnarray}\label{predequno}
\dot{y}_j^S&=&-p y_j^S\sum_{i'}x_{i'}^F +\sum_{i'}q_j P_{i'j}+(1-\theta)\sum_{i'}s_{i'}P_{i'j}+\frac{1}{h}y_j^H,\\\label{predeqdue}
\dot{y}_j^H&=&\nu p y_j^S\sum_{i'}x_{i'}^F+\theta \sum_{i'} s_{i'}P_{i'j} -\frac{1}{h}y_j^H,\\\label{preyequno}
\dot{x}_i^F&=&-px_i^F\sum_{j'}y_{j'}^S+\sum_{j'}q_{j'}P_{ij'}+(1-\theta)\sum_{j'} s_i P_{ij'},\\\label{paireq}
\dot{P}_{ij}&=&(1-\nu)px_i^Fy_j^S-(q_j+s_i)P_{ij},
\end{eqnarray}
with conservation laws for the total predator densities 

\begin{equation}\label{conslawpred}
y_j=y_j^S+y_j^H+\sum_{i'}P_{i'j}
\end{equation}
and for the total prey densities 

\begin{equation}\label{conslawprey}
x_i=x_i^F+\sum_{j'}P_{ij'}.
\end{equation}

Since birth and death are slow processes, on the fast time-scale the total predator densities $y_j$ for each type $j$ are constant, i.e. $\dot{y}_j^S+\dot{y}_j^H+\sum_{i'}\dot{P}_{i'j}=0$. We require the same for the total prey densities $x_i$, i.e.

\begin{equation}
\dot{x}_i=-\nu p x_{i}^F\sum_{j'}y_{j'}^S-\theta \sum_{j'} s_iP_{ij'}=0.
\end{equation}
In order to achieve this, we assume that the predator densities are of a smaller order than the prey densities, i.e. $y_j, y_j^S, y_j^H, P_{ij} \ll x_i, x_i^F$ for all $i,j$, so that in the extreme case $\dot{x}_i=0$ and $x_i^F=x_i$ (see \cite{Berardo:2020aa} and \ref{app3} for details on the time-scale separation method).\\

The functional response $f_{ij}$ of the predator type $j$ for the prey type $i$ is given by the average number of prey type $i$ caught per predator type $j$ per unit of time, i.e. 

\begin{equation}
f_{ij}=\frac{\nu p y_j^S x_i^F+\theta s_i P_{ij}}{y_j},
\end{equation}
with $y_j^S$, $x_i^F$ and $P_{ij}$ at the fast time equilibrium. Therefore, by substituting with the unique equilibrium of the fast dynamics (see (\ref{quasiF}), (\ref{quasiS}) and (\ref{quasiP}) in \ref{app3}), we get

\begin{equation}\label{fr}
f_{ij}=\frac{\nu p x_i+\theta s_i (1-\nu)\frac{p}{q_j+s_i}x_i}{1+hp\sum_{i'}x_{i'}+p (\nu-1) \left[(hq_j-1) \sum_{i'} \frac{x_{i'}}{q_j+s_{i'}} + h(1-\theta) \sum_{i'} \frac{s_{i'}x_{i'}}{q_j+s_{i'}} \right]}.
\end{equation}
We can rewrite the functional response in (\ref{fr}) as 

\begin{equation}
f_{ij}=\frac{\beta_{ij}x_i}{1+\sum_{i'}\beta_{i'j}h_{i'j}x_{i'}},
\end{equation}
with effective capture rate $\beta_{ij}$ and effective handling time $h_{ij}$ defined as

\begin{flalign}\label{effbeta}
\beta_{ij}&=\frac{p}{q_j+s_i}[\nu q_j+s_i(\theta + \nu -\theta\nu)], \\ \label{effh}
h_{ij}&=h+\frac{1-\nu}{\nu q_j+s_i (\theta+\nu-\theta\nu)}.
\end{flalign}

If only one prey type $x$ with strategy $s$ and one predator type $y$ with strategy $q$ are present, then the functional response becomes

\begin{equation}\label{fronetype}
f_{s,q}(x)=\frac{\beta_{s,q} x}{1+\beta_{s,q} h_{s,q}x}
\end{equation}
now written with an explicit argument $x$ of the total prey density, and where 

\begin{flalign}
\beta_{s,q}&=\frac{p}{q+s}[\nu q+s(\theta + \nu -\theta\nu)],\\
h_{s,q}&=h+\frac{1-\nu}{\nu q+s(\theta + \nu -\theta\nu)}
\end{flalign}
are the effective capture rate and handling time, respectively.\\

The functional response in (\ref{fronetype}) is a Holling type II functional response. The average time spent handling after the capture is $h$ and the predator enters the handling state in two cases: with probability $p\frac{q}{q+s}\nu$, a predator gives up stalking the prey and the same prey is captured after a new encounter; with probability $p\frac{s}{q+s}\left(\theta+\nu-\theta\nu\right)$, the prey gives up hiding from the predator and is caught either after escaping or after a new encounter with a predator.
These probabilities define the capture rate $\beta_{s,q}$ in (\ref{fronetype}). 
On the other hand, the encounter and no capture after stalking happens with probability $p(1-\nu)$ and the time spent in the stand-off by the pair is $\frac{1}{q+s}$, therefore, when we multiply the capture rate $\beta_{s,q}$ by the second term $\frac{1-\nu}{\nu q+s(\theta + \nu -\theta\nu)}$ in the handling time $h_{s,q}$, we get $\frac{p(1-\nu)}{q+s}$. \\

Note that when we take the limit for $\nu$ to $1$, the prey and predator never enter the stand-off state and the predator attacks are successful with rate $p$. When this is the case, we obtain the classical version of the Holling type II functional response 

\begin{equation}
f(x)=\frac{px}{1+phx}.
\end{equation}

For the total population dynamics on the slow time-scale we consider a multi-type version of the Rosenzweig-MacArthur model and functional response $f_{ij}$ in (\ref{fr}). The equations for the prey of type $i$ and strategy $s_i$ and the predators of type $j$ and strategy $q_j$ are

\begin{eqnarray}
\dot{x}_i&=&r x_i \left(1-\frac{\sum_{i'} x_{i'}}{K}\right)-\sum_{j'}f_{ij'}(x)y_{j'},\\
\dot{y}_j&=&\gamma\sum_{i'} f_{i'j}(x)y_j - dy_j.
\end{eqnarray}

We use the dimensionless quantities $\tilde{x}_i=\frac{x_i}{K}$, $\tilde{y}_j=\frac{ y_j}{K}$, $\tilde{t}=rt$, $\tilde{\gamma}=\frac{ K\gamma}{r}$, $\tilde{d}=\frac{d}{r}$, $\tilde{\beta}_{ij}=\frac{K\beta_{ij}}{r}$, $\tilde{h}_{ij}=r h_{ij}$ and drop the tildes to obtain
\begin{eqnarray}\label{ecodynx}
\dot{x}_i&=&x_i(1-\sum_{i'} x_{i'})-\sum_{j'}\frac{\beta_{ij'}x_i}{1+\sum_{i'}\beta_{i'j'}h_{i'j'}x_{i'}}y_{j'},\\ \label{ecodyny}
\dot{y}_j&=&\gamma \sum_{i'} \frac{\beta_{i'j}x_i'}{1+\sum_{i'}\beta_{i'j}h_{i'j}x_{i'}}y_j-dy_j.
\end{eqnarray}

\begin{figure}[H]
\centering
\includegraphics[width=.9\textwidth]{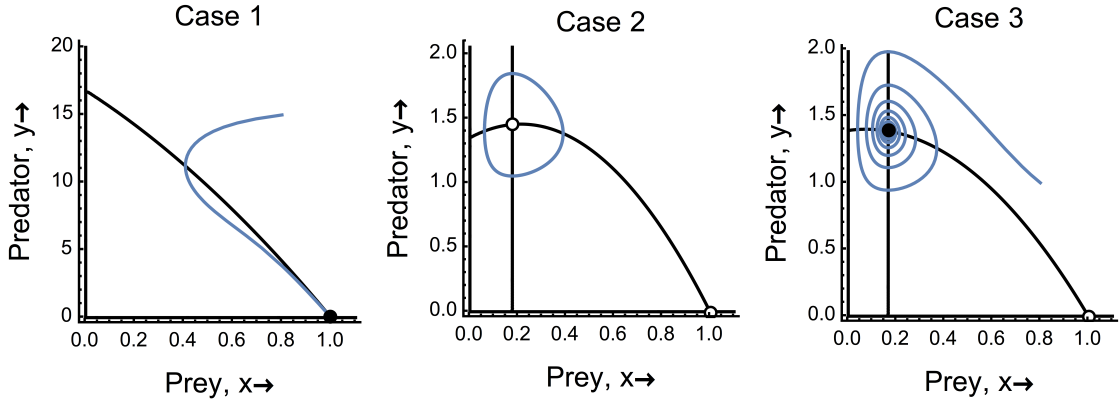}
\caption{Phase plane analysis. Case 1: the interior equilibrium is non-positive and the system converges to the predator-free equilibrium; Case 2: the interior equilibrium is positive but unstable and the system converges to a stable limit cycle, while the predator-free equilibrium is unstable; Case 3: convergence to the positive and stable interior equilibrium, while the predator-free equilibrium is unstable.}
\label{fig:sec1}
\end{figure}

When only a single predator with strategy $q$ and a single prey with strategy $s$ are present, the system is multi-species \cite{rosenzweig1963graphical} model and the outcomes for the ecological dynamics are well-known (see \ref{app2} for details on the bifurcation analysis). The interior equilibrium

\begin{equation}\label{ecoeq}
(\hat{x},\hat{y})=\left(\frac{d}{\gamma \beta_{sq}-d\beta_{sq} h_{sq}},\frac{\gamma (\gamma\beta_{sq} -d \beta_{sq} h_{sq}-d)}{\beta_{sq}^2 (\gamma -dh_{sq})^2}\right)
\end{equation}
is unique and positive if 
 
 \begin{equation}\label{viable}
 \gamma-d h_{sq}>0, \quad \beta_{sq} (\gamma-dh_{sq})>d.
\end{equation} 
The steady state is asymptotically stable if the slope of the prey zero-growth isocline is negative (Figure \ref{fig:sec1}, Case 3), that is if 

\begin{equation}\label{stable}
\gamma+d h_{sq}^2\beta_{sq}+h_{sq}(d-\gamma\beta_{sq})>0.
\end{equation}
When the interior equilibrium changes its stability, a Hopf bifurcation occurs and the system converges to a stable limit cycle (Figure \ref{fig:sec1}, Case 2). In case of non-viability of the interior equilibrium, the predator-free equilibrium is the global attractor of the ecological dynamics (Figure \ref{fig:sec1}, Case 1).

\begin{table}[h]
   \caption{List of model variables, parameters and functions} 
   \label{tab:param}
   \small \centering
   \begin{tabular}{m{2cm} m{8cm}} 
   \textbf{Symbol} & \textbf{Description}  \\ \hline \hline
   $x_i^F$ & foraging prey type $i$\\
   $y_j^S$ & searching predators type $j$\\
   $y_j^H$ & handling predators type $j$\\
   $P_{ij}$ & pair of predator type $j$ and prey type $i$\\
   $x_i$ & total prey density type $i$\\
   $y_i$ & total predator density type $j$\\
   $p$ & encounter rate of predator and prey\\
   $s_i$ & giving-up rate of prey type $i$\\
   $q_j$ & giving-up rate of predator type $j$\\
   $\nu\in [0,1]$ & probability to capture the foraging prey\\
   $\theta\in[0,1] $ & probability to capture the prey when it gives up\\
   $r$ & prey growth rate\\
   $d$ & predator death rate\\
   $h$ & time spent handling the prey\\
   $f_{ij}(x)$ & multi-type predator functional response 
   \end{tabular}
\end{table}

\section{Adaptive dynamics}\label{sec3}
We investigate the coevolution of the rates $s$ and $q$ at which respectively the prey and the predator quit the stand-off state $P$. The strategy of both the prey and the predator is a giving-up rate, which is a continuous trait that could take any non-negative value. We use the mathematical framework of \emph{adaptive dynamics} to understand how the traits evolve and whether coexistence of multiple strategies is favoured by natural selection. We refer to \cite{metz1992should}, \cite{geritz1997dynamics} and  \cite{geritz1998evolutionarily} for the definitions of \emph{invasion fitness} and \emph{selection gradient}.\\

We rewrite the ecological dynamics in equations (\ref{ecodynx}) and (\ref{ecodyny}) in terms of the resident environment $E$

\begin{eqnarray}
\dot{x}_i&=G_{prey}(s_i,q_j,E)x_i,\\
\dot{y}_j&=G_{pred}(s_i,q_j,E)y_j,
\end{eqnarray}
with $E=(e_1,...,e_m,e_1^{prey},...,e_m^{prey},e_1^{pred},...,e_n^{pred})$ and 

\begin{flalign}
e_i^{prey}&=\frac{x_i}{1+\sum_{i'}\beta_{i'j}h_{i'j}x_{i'}},\\
e_j^{pred}&=\frac{y_j}{1+\sum_{i'}\beta_{i'j}h_{i'j}x_{i'}},\\
e_i&=x_i.
\end{flalign}
The instantaneous \emph{per capita} growth rates  $G_{prey}$ and $G_{pred}$ of respectively the prey and the predator are 

\begin{eqnarray}
G_{prey}(s_i,q_j,E)&=&(1-\sum_{i'} e_{i'})-\sum_{j'}\beta_{ij'}e_{j'}^{pred},\\
G_{pred}(s_i,q_j,E)&=&\gamma \sum_{i'} \beta_{i'j} e_{i'}^{prey}-d.
\end{eqnarray}

When the environment settled by the monomorphic resident types with strategies $s$ and $q$ is at the equilibrium, the invasion fitness of a mutant prey with strategy $s_m$ is given by the long-term average population growth rate 

\begin{equation}
g_{prey}(s_m,s,q)=\lim_{t\rightarrow \infty}\frac{1}{t} \int_0^{\infty} G_{prey}(s_m,q,E(t)) dt.
\end{equation}
Similarly, when a mutant predator type with strategy $q_m$ invades the resident dynamics, its invasion fitness is defined by

\begin{equation}
g_{pred}(q_m,s,q)=\lim_{t\rightarrow \infty}\frac{1}{t} \int_0^{\infty} G_{pred}(s,q_m,E(t)) dt.
\end{equation}
The sign of the \emph{invasion fitness} decides whether or not a mutant strategy can invade the resident environment and, in case of convergence to the interior equilibrium, the growth rates of the predator and the prey fully determine the outcome of the invasion. By definition, the invasion fitness verifies $g_{prey}(s,s,q)=0$ and $g_{pred}(q,s,q)=0$ at the ecological equilibrium. Otherwise, when the resident dynamics converges to the periodic attractor, we time-average the population growth rate over the length of the limit cycle with period  $t_{s,q}$

\begin{eqnarray}\label{fitone}
g_{prey}(s_m,s,q)=\frac{1}{t_{s,q}} \int_0^{t_{s,q}} G_{prey}(s_m,q,E(t)) dt,\\\label{fittwo}
g_{pred}(q_m,s,q)=\frac{1}{t_{s,q}} \int_0^{t_{s,q}} G_{pred}(s,q_m,E(t)) dt.
\end{eqnarray}

The direction of the evolution is defined by the sign of the \emph{selection gradient}, i.e. the fitness derivative with respect to the mutant trait and evaluated at the resident strategy. Here, we derivate the fitness of the prey with respect to the mutant strategy $s_m$ and similarly for the predator. The final result for the selection gradient of respectively the prey and the predator is

\begin{flalign}\label{gradx}
\frac{\partial g_{prey}(s_m,s,q)}{\partial s_m}\Big|_{s_m=s}\\\label{grady}
\frac{\partial g_{pred}(q_m,s,q)}{\partial q_m}\Big|_{q_m=q}
\end{flalign}
Analogously to the invasion fitness, the selection gradient in a cycling resident population is defined by the time-average of the expressions in (\ref{gradx}) and (\ref{grady}) over the length of the limit cycle.\\

A \emph{singular strategy} is a pair of values $(s^*,q^*)$ for the coevolving strategies $s$ and $q$ such that it is an intersection point for the isoclines

\begin{equation}
\frac{\partial g_{prey}(s_m,s,q)}{\partial s_m}\Big|_{s_m=s=s^*,q=q^*}=0, \quad \frac{\partial g_{pred}(q_m,s,q)}{\partial q_m}\Big|_{q_m=q=q^*,s=s^*}=0.
\end{equation}
When the singularity $(s^*,q^*)$ is a local maximum of the invasion fitness, that is the second derivatives 

\begin{equation}
\frac{\partial^2  g_{prey}(s_m,s,q)}{\partial s_m^2}\Big|_{s_m=s=s^*,q=q^*}<0,\quad \frac{\partial^2  g_{pred}(s_m,s,q)}{\partial q_m^2}\Big|_{q_m=q=q^*,s=s^*}<0,
\end{equation}

then the evolutionary singular strategy cannot be invaded by any mutant prey or predator trait (see the definition of ESS by \cite{smith1982evolution}, later extended to asymmetric games and coevolutionary ESS, for example see \cite{vincent1988evolution}). Conditions for $(s^*,q^*)$ to be an evolutionary attractor are given in \ref{app1}.\\

A \emph{boundary attractor} $(s^*,0)$ satisfies 

\begin{equation}
\frac{\partial g_{prey}(s_m,s,q)}{\partial s_m}\Big|_{s_m=s=s^*,q=q^*}=0, \quad \frac{\partial g_{pred}(q_m,s,q)}{\partial q_m}\Big|_{q_m=q=q^*,s=s^*}<0,
\end{equation}
and similarly for $(0,q^*)$. \\

The \emph{canonical equation} of adaptive dynamics (see \cite{dieckmann1996dynamical},  \cite{Champagnat01arous}) describes the rate of change of the traits $s$ for the prey and $q$ for the predator with the following relations

\begin{eqnarray}\label{evodyns}
\dot{s}&=k_{prey}(s,q) \frac{\partial g_{prey}(s_m,q)}{\partial s_m}\big|_{s_m=s},\\\label{evodynq}
\dot{q}&=k_{pred}(s,q) \frac{\partial g_{pred}(s,q_m)}{\partial q_m}\big|_{q_m=q}
\end{eqnarray}
where $k_{prey}(s,q)$ and $k_{pred}(s,q)$ are scaling non-negative coefficients rather difficult to derive and which take into account the \emph{influence of mutation}. In particular, $k_{prey}(s,q)=\frac{1}{2}\mu_{prey}(s)\sigma_{prey}^2(s)x_E(s,q)$, where $\mu_{prey}(s)$ is the mutation probability per birth event, $\sigma_{prey}^2(s)$ is the variance of the mutation step distribution and $x_E(s,q)$ the effective population size (the resident equilibrium population size if we assume the resident population at equilibrium). The same definition holds for $k_{pred}(s,q)=\frac{1}{2}\mu_{pred}(q)\sigma_{pred}^2(q)y_E(s,q)$. \\

We refer to the works by \cite{ripa2013mutant} and \cite{metz2016canonical} for the extension of the canonical equation to a periodic environment. In particular, the drifts given by \cite{ripa2013mutant} differ of a factor $\frac{1}{2}$ from the original canonical equations by \cite{dieckmann1996dynamical}, as this is embedded into the definitions for the effective population sizes: the coefficient for the prey trait equation is now $k_{prey}(s,q)=\mu_{prey}(s)\sigma_{prey}^2(s)x_E(s,q)$, and similarly for the predator trait, $k_{pred}(s,q)=\mu_{pred}(q)\sigma_{pred}^2(q)y_E(s,q)$. \\

In the expressions for $k_{prey}(s,q)$ and $k_{pred}(s,q)$, the terms $\mu_{prey}(s)$ and $\sigma_{prey}^2(s)$, $\mu_{pred}(q)$ and $\sigma_{pred}^2(q)$ are additional model parameters and do not follow from the population dynamics. Conversely, we define the effective population densities with the following ratios between time averages over one complete population cycle

\begin{eqnarray}\label{effx}
x_E(s,q)&=\frac{ \int_0^{t_{s,q}} (b_{prey}+d_{prey}) dt}{2  \int_0^{t_{s,q}} \frac{b_{prey}+d_{prey}}{x_{s,q}} dt},\\\label{effy}
y_E(s,q)&=\frac{ \int_0^{t_{s,q}} (b_{pred}+d_{pred}) dt}{2  \int_0^{t_{s,q}} \frac{b_{pred}+d_{pred}}{y_{s,q}} dt}.
\end{eqnarray}
The terms $b_{prey}$ and $b_{pred}$ are the explicit birth terms in the prey and predator equations, while $d_{prey}$ and $d_{pred}$ refer to the corresponding death terms. In the same way, the selection gradients in equations (\ref{evodyns}) and (\ref{evodynq}) are averaged over the length of the limit cycle.

\section{Predator-prey coevolution}\label{sec4}

\subsection{Evolutionary dynamics: invasion fitness and selection gradient}\label{sec41}
We assume that the resident prey and predator populations are monomorphic most of the time. When an invasion occurs, the mutant population is sufficiently rare compared to the resident species and we can apply time-scale separation between the ecological dynamics on the fast time-scale and the evolutionary dynamics on the slow time-scale. Therefore, we assume that the resident population has attained an ecologically stable attractor when the mutant comes along. After invasion, the population evolves towards an evolutionary attractor through a sequence of trait substitutions. During the process of directional evolution, the population remains monomorphic except for the infinitely short time straight after invasion. \\

We suppose the monomorphic prey population with strategy $s$ and the monomorphic predator population with strategy $q$. The analytical results collected below are displayed in the $(s,q)$-planes of Figure \ref{fig:sec21} and Figure \ref{fig:sec22}.\\

The invasion fitness of a mutant prey with strategy $s_m$ in the constant environment defined by the resident populations is

\begin{equation}\label{monofitx}
g_{prey}(s_m,s,q)=(1-\hat{x})-\frac{\beta_{s_m,q} \hat{y}}{1+\beta_{s,q}h_{s,q}\hat{x}},
\end{equation}
with $(\hat{x},\hat{y})$ as defined in (\ref{ecoeq}), $\beta_{s,q}$ and $h_{s,q}$ as defined for (\ref{fronetype}) and

\begin{equation}\label{smutdepbeta}
\beta_{s_m,q} =\frac{p}{q+s_m}[\nu q+s_m (\theta+\nu-\theta\nu)].
\end{equation}
Note that $\beta_{s_m,q}$ is the only factor depending on the mutant trait $s_m$, as the ratio $\frac{\hat{y}}{1+\beta_{s,q}h_{s,q}\hat{x}}$ represents the fraction of resident searching predators to which the mutant prey is subjected.\\

Similarly, the invasion fitness for a mutant predator with strategy $q_m$ is given by

\begin{equation}\label{monofity}
g_{pred}(q_m,s,q)=\gamma \frac{\beta_{s,q_m} \hat{x}}{1+\beta_{s,q_m} h_{s,q_m} \hat{x}}-d,
\end{equation}
with $\hat{x}$ as defined in (\ref{ecoeq}) and 

\begin{eqnarray}\label{qmutdepbeta}
\beta_{s,q_m}&=&\frac{p}{q_m+s}[\nu q_m+s (\theta+\nu-\theta\nu)],\\ \label{qmutdeph}
h_{s,q_m}&=&\frac{1-\nu}{\nu q_m+s(\theta + \nu -\theta\nu)}+h.
\end{eqnarray}

The evolutionary change is determined by the selection gradient. The selection gradient for the values of trait $s$ is given by the derivative of the invasion fitness (\ref{monofitx}) with respect to $s_m$ evaluated at $s=s_m$,

\begin{equation}\label{monogradx}
\frac{\partial g_{prey}(s_m,s,q)}{\partial s_m}\Big|_{s_m=s}=\frac{ \hat{y} \frac{\partial\beta_{s_m,q}}{\partial s_m}\big|_{s_m=s}}{1+\beta_{s,q}h_{s,q}\hat{x}}=\frac{-pq\theta(1-\nu)}{(q+s)^2}\frac{\hat{y}}{1+\beta_{s,q}h_{s,q}\hat{x}}.
\end{equation}
In the same way, we define the gradient with respect to $q_m$ as

\begin{equation}\label{monogrady}
\frac{\partial g_{pred}(q_m,s,q)}{\partial q_m}\Big|_{q_m=q}= \frac{p\gamma(1-\nu) (p\nu \hat{x}-s\theta)}{[q+s+p\hat{x}(1-\nu+hq\nu+hs(\theta+\nu-\theta\nu))]^2}.
\end{equation}
The unique predator isocline follows from $dg_{pred}(s,q)=0$ and is the vertical line

\begin{equation}\label{prediso}
s=\frac{d}{(\gamma-dh)\theta}.
\end{equation}
By imposing $dg_{prey}(s,q)=0$, we obtain the prey isoclines 

\begin{flalign}\label{preyisoone}
&q_1=0,\\ \label{preyisotwo}
&q_2=\frac{d p [\nu -1-h s ( \theta+\nu-\theta\nu )]-d s+\gamma  p s (\theta +\nu- \theta\nu)}{d h \nu  p+d-\gamma  \nu  p}.
\end{flalign}
Note that the prey isocline $q_2$ is a straight line with slope $\frac{p(\gamma -dh) ( \theta+\nu-\theta\nu )-d}{d h \nu  p+d-\gamma  \nu  p}$ and intersects the horizontal axis at $s=\frac{d (1-\nu) p}{p(\gamma -dh) ( \theta+\nu-\theta\nu )-d}$. It coincides with the \emph{extinction boundary} for the ecological dynamics where a transcritical bifurcation occurs and the interior equilibrium interchanges its stability with the predator-free equilibrium (see \ref{app2}). 
Furthermore, when $\gamma > dh$ and $p=\frac{d}{(\gamma-dh)\nu}$, the prey isocline coincides with the vertical predator isocline. When $\gamma > dh$ and $\frac{d}{(\gamma-dh)(\theta+\nu-\theta\nu) }<p<\frac{d}{(\gamma-dh)\nu}$, the slope of the prey isocline is positive and viceversa for $p>\frac{d}{(\gamma-dh)\nu}$. The same conditions can be given also in terms of the parameters $\nu$ and $\theta$.\\

The intersection of the isoclines (\ref{prediso}) and (\ref{preyisoone}) gives the unique singular point $(s^*,q^*)=\left(\frac{d}{(\gamma-dh)\theta},0\right)$ (with respect to both traits $s$ and $q$) when the ecological dynamics attains its interior equilibrium. In particular, $s^*$ is positive if and only if $\gamma>dh$. \\

As $\frac{\hat{y}}{1+\beta_{s,q}h_{s,q}\hat{x}}>0$, the sign of prey gradient is only determined by the ratio $\frac{-pq\theta(1-\nu)}{(q+s)^2}$ in (\ref{monogradx}), that is always negative for $\nu\in[0,1)$. Similarly, we study the sign of the predator gradient which depends only on the factor $(p\nu \hat{x} - s\theta)$ in (\ref{monogrady}), since $1-\nu>0$ and the denominator is always positive. Therefore, the predator gradient verifies

\begin{eqnarray}\label{gradcond1}
\frac{\partial g_{pred}(q_m,s,q)}{\partial q_m}\Big|_{q_m=q}>0 \quad \text{if} \quad \hat{x}>\frac{s\theta}{p\nu},\\\label{gradcond2}
\frac{\partial g_{pred}(q_m,s,q)}{\partial q_m}\Big|_{q_m=q}=0 \quad \text{if} \quad \hat{x}=\frac{s\theta}{p\nu},\\\label{gradcond3}
\frac{\partial g_{pred}(q_m,s,q)}{\partial q_m}\Big|_{q_m=q}<0 \quad \text{if} \quad \hat{x}<\frac{s\theta}{p\nu}.
\end{eqnarray}
Thus, high giving-up rates (and shorter stand-offs) for the predator are advantageous if the prey is abundant, and, viceversa, low giving-up rates (and longer stand-offs) are better when the prey is rare. In other words, when the prey density is above the critical value $\frac{s\theta}{p\nu}$, finding a new prey is fast relative to waiting out the stand-off, which, in this scenario, is a waste of time for the predator.\\

The above conditions on the sign of the predator gradient can be reformulated in terms of the singularity $s^*$. We check the sign of $(p\nu \hat{x} - s\theta)$ with full expression for $\hat{x}$ and $q=0$. We solve for $s$ and obtain that the predator giving-up rate switches downward evolution to upward evolution, creating a \emph{bang-bang situation}, once the prey strategy passes the singular point $s^*$. The singularity $(s^*,0)$ is an unstable saddle and the conditions for convergence stability do not apply here as well as we can exclude evolutionary branching. Furthermore, the points  $(s, 0)$ with $s>s^*$ are (non-isolated) \emph{boundary attractors} with respect to $s$, as they lay on the prey isocline $q_1$. \\

In \ref{app4} we prove the main result on the predator gradient when the resident dynamics is at the interior equilibrium, i.e. the biological explanation for the directional evolution of the predator trait. We summarise this with the following Proposition 
\begin{prop}\label{prop1}
In a constant resident population, when $s=s^*$, with $(s^*,0)$ being the unique singularity, then the expected time till prey capture for a predator starting in the searching state ($ES$) is equivalent to the expected time till prey capture for a predator starting in the stand-off state ($EP$). If $s>s^*$, then $ES>EP$. Finally, if $s<s^*$, then $ES<EP$.
\end{prop}

When the population is cycling, we use standard numerical methods to compute the invasion fitness and selection gradient. Specifically we use the procedure \emph{NDSolve} of the software \emph{Mathematica}$^\circledR$ to numerically integrate the population equations in (19) and (20) for one prey and one predator. We use the Poincar\'e map to evaluate convergence of the solutions. In particular, we collect the data until the distance between $x$ and $R(x)$, with $R$ denoting the return map, is less than a small error tolerance. The length of the limit cycle is measured by the time interval between $x$ and $R(x)$. Finally we numerically integrate the mutant's growth rate over the limit cycle as indicated in equations (\ref{fitone}) and (\ref{fittwo}).\\

We find 10 different configurations for the evolutionary phase plane with respect to the giving-up rates $s$ and $q$, which are given both in Figure \ref{fig:sec21} and Figure \ref{fig:sec22}. In Scenario 1-6 we vary the parameter $\nu\in[0,1)$. We distinguish the areas where the interior equilibrium of the ecological dynamics is non-positive (and the resident population attains the predator-free equilibrium), or positive and stable, or positive and unstable (and the resident population converges to a stable limit cycle). Note that the extinction boundary never coincides with the Hopf bifurcation line (see the top-right panel which zooms around the Hopf and transcritical bifurcations and \ref{app2}). Finally, we give the prey isoclines, the predator isoclines and the unique singularity. By continuity, the vertical isocline extends to the region of convergence to the stable limit cycle in Scenario 2 and 3, as well as the prey isocline $q_1=0$ is still present in the cycling region. Once in the cycling region, the predator isocline is no longer vertical, but its slope is determined by the periodic resident population. \\

We fix $\nu=0.5$ and we obtain Scenario 7 and 8 by varying the parameter $\theta\in(0,1]$. In particular, when $\theta=0$, the predator isocline in (\ref{prediso}) is at infinity, while it becomes feasible for parameter values greater than $0$ (Scenario 8). In Scenario 7, we show that the predator isocline can also fall on the right hand side of the cycling region for small values of $\theta$.\\

In the same way, we fix $\nu=0.01$ and we check the dynamics for different values of $\theta\in(0,1]$. We add Scenario 9 and 10 to the list of possible evolutionary phase planes: the prey isocline has positive slope and the predator isocline appears in the region of non-viability for the interior equilibrium. In Scenario 10 there are no values for $(s,q)$ such that the interior equilibrium is positive and, therefore, the dynamics only converges to the predator-free equilibrium. \\

Note that if $\nu=1$ and the stand-off never occurs, we obtain the degenerate case where the prey and the predator gradients in (\ref{monogradx}) and (\ref{monogrady}) are zero for every value of $s$ and $q$. On the other hand, when $\theta=0$ and the stand-off never ends with prey capture, the prey gradient is zero for every $s$ and $q$, while the predator gradient is positive everywhere.

\subsection{Evolutionary dynamics: the canonical equation of adaptive dynamics}\label{sec42}
We use the canonical equations in (\ref{evodyns}) and (\ref{evodynq}) and give the stream plots of the prey drift $k_{prey}(s,q) \frac{\partial g_{prey}(s_m,q)}{\partial s_m}\big|_{s_m=s}$ and the predator drift $k_{pred}(s,q) \frac{\partial g_{pred}(s,q_m)}{\partial q_m}\big|_{q_m=q}$ to understand the direction of the evolution of the traits $s$ and $q$.
In particular, we fix the mutation probability per birth event $\mu_{prey}(s)$ and $\mu_{pred}(q)$, and the mutation variance $\sigma_{prey}^2(s)$ and $\sigma_{pred}^2(q)$ (note that for the shape of the orbits only the relative values matter). \\

In the cycling region of the $(s,q)$-plane, we plot the drifts of the canonical equations in (\ref{evodyns}) and (\ref{evodynq}) with effective population sizes in (\ref{effx}) and (\ref{effy}). In particular, this definition forces us to split the right-hand side of the population equations into explicit birth and death terms. How to do this is trivial, except for the prey growth term $x(1-x)$. One possibility is $x(1-x)=2x-x(1+x)$, with $2x$ being the birth term and $x(1+x)$ modelling (density-independent) natural mortality and (density-dependent) death due to intraspecific competition between the prey individuals. Therefore the birth and death \emph{per capita} rates that we choose for the numerical analysis are

\begin{flalign}
&b_{prey}=2\\
&d_{prey}=1+x+\frac{f_{s,q}(x)}{x}y\\
&b_{pred}=\gamma f_{s,q}(x)\\
&d_{pred}=d.
\end{flalign}

\begin{figure}[H]
\captionsetup[subfigure]{labelformat=empty}
\centerline{\begin{subfigure}[b]{1.3\textwidth}
\includegraphics[width=\textwidth]{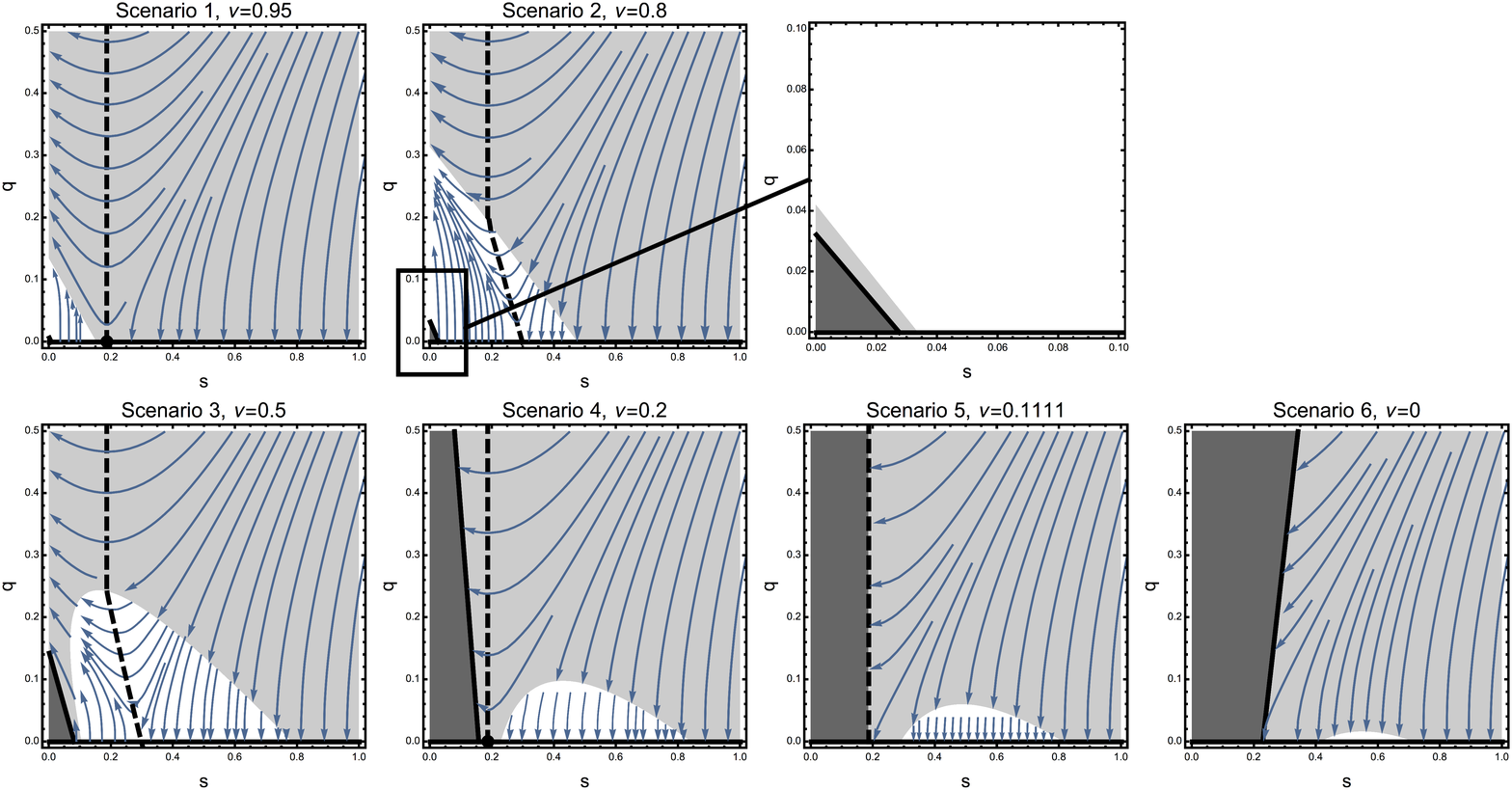}
  \caption{Parameter values for Scenario 1-6: $d=0.1,\gamma =1,h=1,p=1,\theta =0.6.$}
        \label{fig:21a}
\end{subfigure}}
\par\bigskip
\centerline{\begin{subfigure}[b]{1.3\textwidth}
\includegraphics[width=\textwidth]{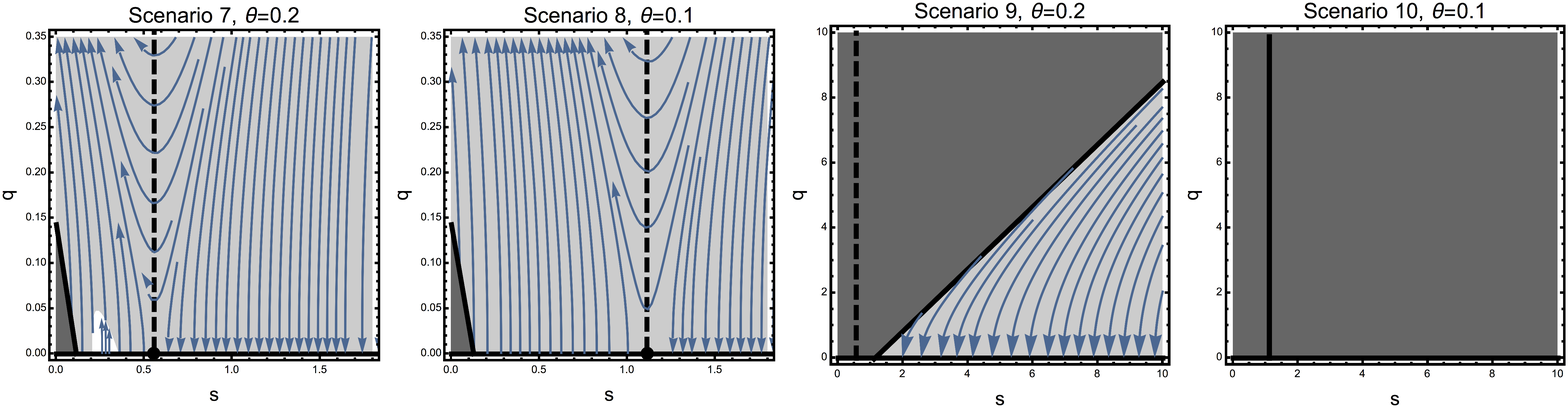}
  \caption{Parameter values for Scenario 7 and 8: $d=0.1,\gamma =1,h=1,p=1,\nu =0.5$. Parameter values for Scenario 9: $d=0.1,\gamma =1,h=1,p=1,\nu =0.01.$ }
        \label{fig:21b}
\end{subfigure}}
\caption{Evolutionary phase planes for $\nu\in[0,1]$ and $\theta\in(0,1]$. Dark grey region: non-viability for the resident interior equilibrium. Light grey region: viability and asymptotic stability for the resident interior equilibrium. White region: unstable resident interior equilibrium and stable limit cycle. Black lines: prey isoclines. Dotted line: predator isocline. Thick point: the unique singularity. Blue arrows: orbits of the system of canonical equations ($\mu_{prey}(s)=1, \sigma_{prey}^2(s)=1, \mu_{pred}(q)=1, \sigma_{pred}^2(q)=1$).}
\label{fig:sec21}
\end{figure}

\begin{figure}[H]
\captionsetup[subfigure]{labelformat=empty}
\centerline{\begin{subfigure}[b]{1.3\textwidth}
\includegraphics[width=\textwidth]{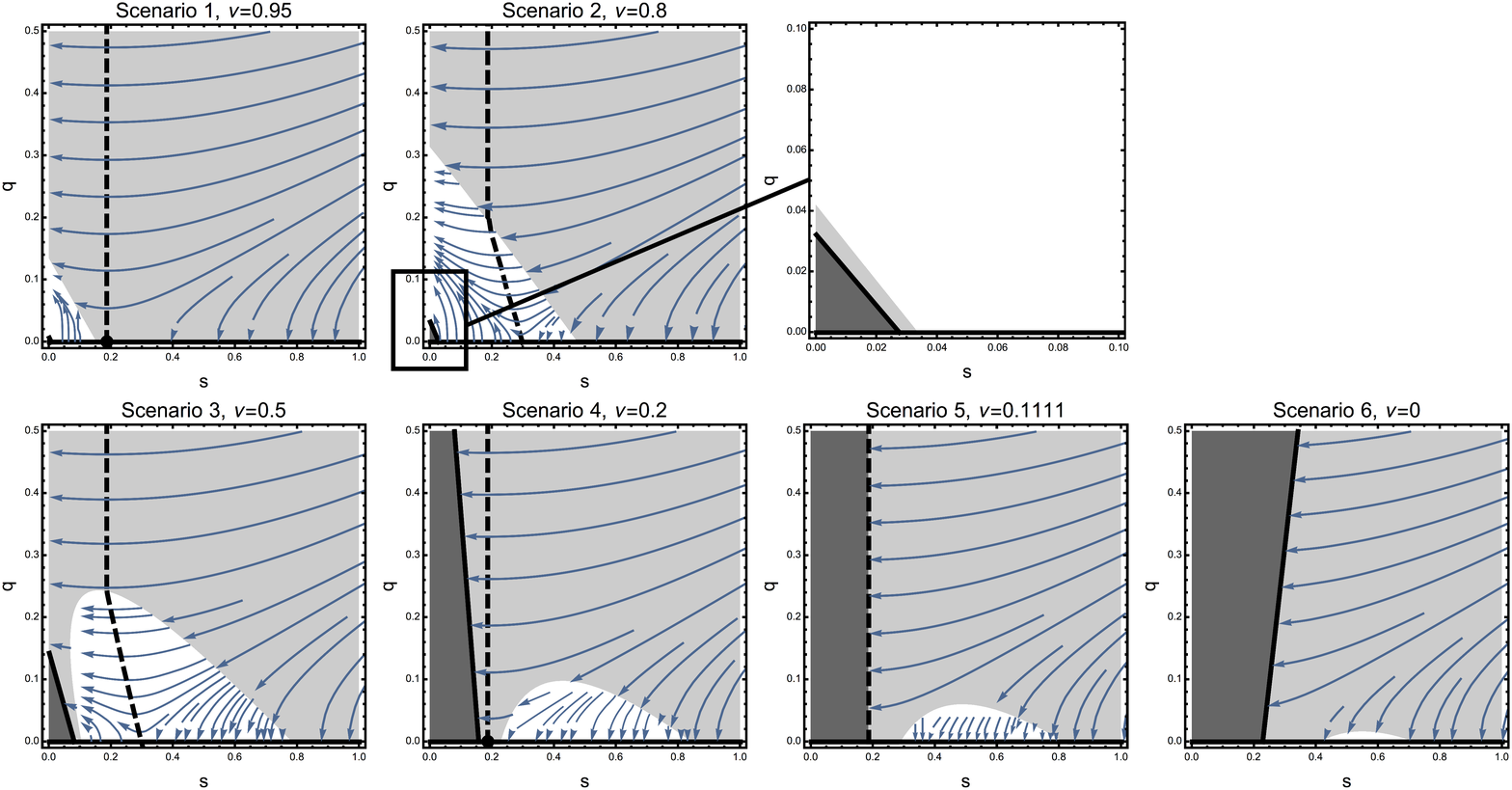}
  \caption{Parameter values for Scenario 1-6: $d=0.1,\gamma =1,h=1,p=1,\theta =0.6.$}
        \label{fig:22a}
\end{subfigure}}
\par\bigskip
\centerline{\begin{subfigure}[b]{1.3\textwidth}
\includegraphics[width=\textwidth]{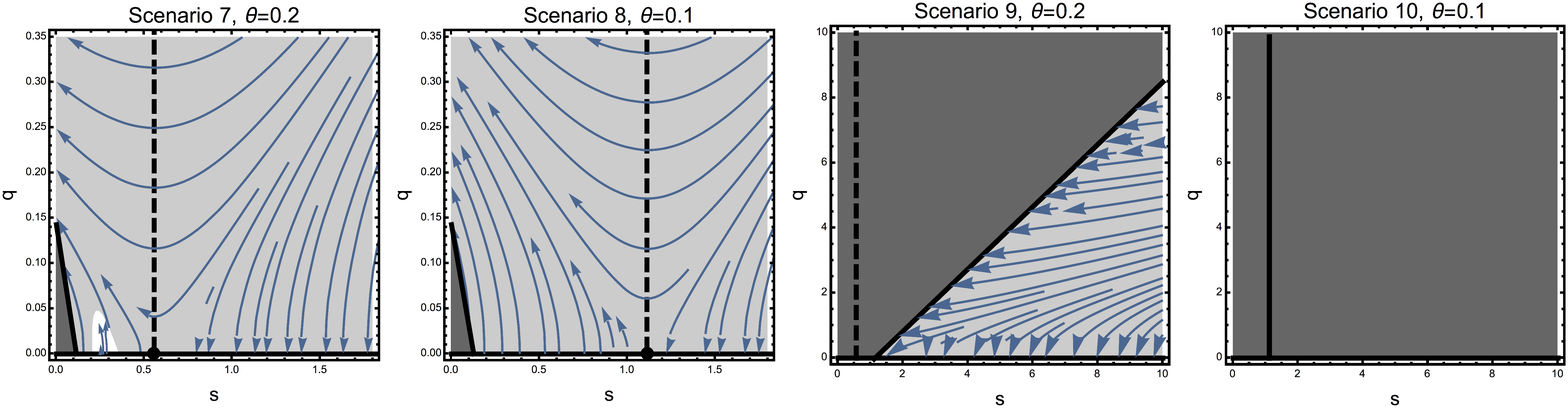}
  \caption{Parameter values for Scenario 7 and 8: $d=0.1,\gamma =1,h=1,p=1,\nu =0.5$. Parameter values for Scenario 9: $d=0.1,\gamma =1,h=1,p=1,\nu =0.01.$ }
        \label{fig:22b}
\end{subfigure}}
\caption{Evolutionary phase planes for $\nu\in[0,1]$ and $\theta\in(0,1]$. Dark grey region: non-viability for the resident interior equilibrium. Light grey region: viability and asymptotic stability for the resident interior equilibrium. White region: unstable resident interior equilibrium and stable limit cycle. Black lines: prey isoclines. Dotted line: predator isocline. Thick point: the unique singularity. Blue arrows: orbits of the system of canonical equations ($\mu_{prey}(s)=1, \sigma_{prey}^2(s)=10, \mu_{pred}(q)=1, \sigma_{pred}^2(q)=1$).}
\label{fig:sec22}
\end{figure}

The resulting dynamics is described in Figure \ref{fig:sec21} and Figure \ref{fig:sec22}, which differ in the speed of evolution for the prey trait, $\sigma_{prey}^2(s)$.\\

Note that in Scenario 10 the dynamics converges only to the predator-free equilibrium and there is no directional evolution. Furthermore, we do not give the case $\nu=1$, where both selection gradients are null and so are the drifts. The case $\theta=0$ is also not displayed: here the prey drift is zero for every $s$ and $q$ (as the prey gradient is zero) and all the trajectories are vertical and converging to high levels of $q$ if the prey isocline $q_2$ has negative slope, some of them converging to the extinction boundary otherwise. \\

In Scenario 1, 7 and 8 in Figure \ref{fig:sec21} the vertical isocline falls on the right-hand side of the cycling region. The pair of traits $(s,q)$ will approach either the boundary attractors with $s>s^*$ and $q=0$, or eventually meet the vertical axis $s=0$ and evolve towards values of $q\gg 0$. In particular, the prey drift is negative everywhere above the prey isocline $q_1$, while the predator drift changes sign when the orbits cross the predator isocline. Note that on the isocline $q_2$ the prey drift is zero, given that the prey gradient $\frac{\partial g_{prey}(s_m,q)}{\partial s_m}\big|_{s_m=s}=0$. At the same time, $\frac{dq}{dt}$ becomes zero in the region of convergence to the predator-free equilibrium where the predator effective population size is $y_E(s,q)=0$.  \\

In Scenario 2 and 3 in Figure \ref{fig:sec21}, the vertical isocline falls into the region of convergence to the stable limit cycle. Some trajectories of the evolutionary dynamics enter the cycling region from the right-hand side of the predator isocline. However, when the population is cycling, the predator drift does not change its behaviour, as it is negative on the right-hand side of the predator isocline, and positive otherwise. \\

Till now we have observed two dynamics, which occur depending on the model parameters and the initial conditions. In terms of quitting times, convergence to the boundary attractors with $s>s^*$ and $q=0$ reads like the predator never gives up, and the prey gives up after an exponentially distributed amount of time. Otherwise, convergence to the vertical axis and high values of $q$ results in the case when the predator gives up immediately, and the prey never gives up.\\

In Scenario 4, 5 and 6 in Figure \ref{fig:sec21} we observe a new behaviour in addition to the ones described above: some trajectories of the evolutionary dynamics will eventually end into the extinction boundary, which is here attracting. As the predator density declines, selection decreases, because of continuity of the fitness gradient as a function of the environment and the traits $s$ and $q$, and, in the absence of the predator, it becomes neutral (i.e. there is no longer directional selection). The orbits which end on the extinction boundary become almost horizontal, as $\frac{dq}{dt}$ becomes very small. The points on the prey isocline are also (non-isolated) attractors for the prey trait $s$ and, as a consequence, the pair of parameter traits stops evolving when it encounters the isocline.  \\

Finally, in Scenario 9 in Figure \ref{fig:sec21}, both the prey and the predator drifts are negative and, as a consequence, directional evolution is always towards the points with $s>s^*$ and $q=0$.\\

As directional selection is determined by the relative speeds of evolution, the extinction boundary is repelling only if its slope is negative enough compared to the one of the stream lines. When we increase the speed of evolution of the prey trait $s$, we observe that convergence to the extinction boundary becomes a more likely outcome, as shown in Figure \ref{fig:sec22}. In other words, a sort of \emph{evolutionary murder} occurs, since evolution in the $s$-direction determines the extinction of the predator species.

\section{Conclusions}\label{sec5}
Using the mathematical framework of adaptive dynamics, we studied the coevolution of the giving-up rates $q$ and $s$ of respectively the predator and the prey in a stand-off situation after a failed attack. We found three qualitatively different long-term evolutionary outcomes depending on the model parameters as well as the initial conditions of the strategy dynamics:
\begin{enumerate}[label=(\roman*)]
   \item\label{outcome1} the predator gives up immediately (i.e., $q=\infty$), while the prey never gives up (i.e., $s=0$);
   \item\label{outcome2} the predator never gives up (i.e., $q=0$), while the prey adopts any giving-up rate greater than or equal to a given positive threshold value;
   \item\label{outcome3} the predator goes extinct.  
\end{enumerate}

Concerning the transient phase of the strategy dynamics, we found:
\begin{enumerate}[resume,label=(\roman*)]
   \item\label{outcome4} the giving-up rate $s$ of the prey always decreases unless the predator has gone extinct or has adopted a giving-up rate $q=0$, in which cases $s$ has become selectively neutral;
   \item\label{outcome5} the giving-up rate $q$ of the predator decreases for high values of the giving-up rate $s$ of the prey, but it increases once $s$ has become less than a given threshold value.
\end{enumerate}
Graphical examples of the various scenarios are given in Figure \ref{fig:sec21} and Figure \ref{fig:sec22}. Note that results \ref{outcome1} and \ref{outcome2} are theoretical; in practice, the prey refuses to give up before the predator gives up in \ref{outcome1}, and similarly for the predator in \ref{outcome2}. Furthermore, these are fast time-scale processes and the population will not go extinct, as birth and death happen on the slower time-scale where even an arbitrarily long time is negligibly short if measured on the slow time-scale.\\

The threshold values in \ref{outcome2} and \ref{outcome5} are the same. For a constant (i.e., non-cycling) population, we have shown (see Proposition \ref{prop1}) that if $s$ is equal to this threshold, denoted by $s^*$, then the expected time till the next successful prey capture for a searching predator and for a predator in a stand-off are exactly the same. For $s>s^*$ the expected time till the next successful prey capture for a searching predator is longer than for a predator in a stand-off, whereas for $s<s^*$ the situation is reversed. Therefore, if $s>s^*$, there is a selective advantage for the individual predator to adopt an even lower giving-up rate, whereas if $s<s^*$, then the advantage is for the individual predator with a higher giving up rate.
In other words, evolution minimises the expected time till the next prey capture, which is equivalent to maximising the rate of prey capture, which again is equivalent to maximising the predator's \emph{per capita} birth rate.\\

The above explains the results \ref{outcome1}, \ref{outcome2} and \ref{outcome5}, at least for a constant population. The same argument holds for a cycling population as well, except that the threshold value of $s$ is no longer a constant but depends on the value of $q$ (see Figure \ref{fig:sec21}, Scenario 1, 2 and 3), and, moreover, that the expected time till the next successful prey capture also involves averaging over the population cycle. Figure \ref{fig:sec21} also shows that \ref{outcome1} is the more likely result if the predator already starts with a high giving-up rate or evolves slower than the prey, while we get \ref{outcome2} more often if the prey starts with a high giving-up rate and the predator evolves faster than the prey.\\

The costs or benefits for the prey have a completely different origin than for the predator. To understand the result \ref{outcome4} we observe from the invasion fitness of the prey in equation (\ref{monofitx}) that the giving-up rate $s_m$ of the mutant prey affects the invasion fitness of the prey through the effective capture rate $\beta_{s_m,q}$ in the numerator of the predator's functional response, while the denominator only depends on the resident strategies. Since $\beta_{s_m,q}$ is an increasing function of $s_m$, it is always beneficial for the individual prey to have a lower giving-up rate. In other words, evolution minimises the predation-related \emph{per capita} death rate of the prey, irrespectively of the resident strategies. \\

The predator goes extinct if the prey capture rate for a searching predator is higher than for a predator in a stand-off. However, as we observe in scenarios 3, 4, 7 and 8 of Figure \ref{fig:sec21}, the predator cannot take advantage of this, because the giving-up rates of both the predator and the prey are low, so that the two stay too long in the stand-off situation. Alternatively, the predator goes extinct if the prey capture rate in the stand-off is higher than when searching, but the predator again cannot take advantage of this. In this case, the giving-up rate of the predator is high, so that the stand-off is left too early (Figure \ref{fig:sec21}, scenarios 3, 4, 7 and 8). Result \ref{outcome3} happens when an evolutionary trajectory crosses the extinction boundary in the $(s,q)$-plane. From the direction of the evolutionary trajectories it can be seen in Figure \ref{fig:sec21} that it is always the evolutionary change in the $s$-direction, not in the $q$-direction, that drives the population over the extinction boundary. It is therefore correct to say that the prey drives the predator to extinction. It also follows that extinction becomes more likely if the prey evolves faster than the predator, so that the horizontal component of the coevolutionary velocity vector becomes dominant (Figure \ref{fig:sec22}).\\

In the context of evolutionary game theory, the asymmetric war of attrition has a continuum of Nash equilibria where one player quits immediately while the other is prepared to wait any time at a cost that is not less than the value of the contested object (\cite{smith1974theory}, \cite{smith1976logic}, \cite{selten1980note}, \cite{kim1993evolutionary}). As a consequence of our model being derived from individual level interactions with exponentially distributed event times, quitting after a fixed positive time cannot be expressed in terms of a constant giving up rate. In that sense, the Nash equilibria of the war of attrition in game theory cannot be reproduced here. At most we can express evolutionary outcomes in terms of average giving-up times, which are the reciprocal of the corresponding giving-up rates. With that in mind, only result \ref{outcome1} is similar to the Nash equilibria of the asymmetric war of attrition.\\

All other results are enigmatic for the present model. Result \ref{outcome2} is the opposite of the Nash equilibrium: one player never gives up (as apposed to immediately giving up), while the other player can adopt any giving-up rate greater than a given threshold rate, which is the same as having an average giving-up time that is less than (as opposed to greater than) a given positive threshold time. Results \ref{outcome3}, \ref{outcome4} and \ref{outcome5} are about (or a direct consequence of) strategy dynamics by selection and small mutation steps and dependence on initial conditions. This is the realm of the adaptive dynamics approach and such results cannot be reproduced by a game theoretical analysis because of its focus on evolutionary stability and Nash equilibria.\\

While our results differ from the asymmetric war of attrition game partly because of the chosen methods of analysis, more important, however, is the difference in modelling approaches. The war of attrition is formulated without ecological context, and the costs and benefits for the players are predetermined given their respective strategies. Embedding the game in a population dynamical model such as the \emph{replicator equation} by \cite{hofbauer2003evolutionary} does not add any new ecology. In contrast, our model is derived from a network of individual-level interactions and processes including the formation and break-up of a predator-prey pair engaged in the mutual stand-off. As a consequence, the evolutionary game is an integral part of the full ecology and the predator-prey dynamics. The costs and benefits of different strategies are implicit and emerge from the dynamics rather than being predetermined. Only after the model was formulated and analysed, it emerged that the costs and benefits for the predator can be measured in terms of the \emph{per capita} birth rate, while for the prey they are measured in terms of the \emph{per capita} death rate due to predation. This may seem obvious, but only so in retrospect, after the model has been formulated and analysed. And as seen above, the results are not the same.\\

Note that the time scale separation is another crucial step in our modelling approach. By the singular perturbation theory, we obtain qualitatively similar results if we relax the assumptions of time scale separation a little bit. By relaxing the time scale separation assumptions even further, the system becomes multi-dimensional and has potentially more complex dynamics, including chaotic behaviour and multiple attractors. In this scenario, we do not know to what extent our conclusions still apply.\\

An interesting and quite natural extension of the model would include the possibility that the stand-off escalates into a fight where both the predator and the prey may get injured. The costs and benefits of different giving-up times then would not only come from changes in the \emph{per capita} birth rates (in case of the predator) and predation-related death rates (in case of the prey) as in the present model, but also from the effects of injury on the fecundity and mortality rates of both the predator and the prey. Although it is not obvious how the interaction between the stand-offs and the escalated fights would affect the evolutionary outcomes, using our modelling approach (from individual-level processes to population dynamics) makes the evolutionary problem easy to formulate and straightforward to analyse with the adaptive dynamics framework.

%


\appendix

\section{On the time-scale separation for the ecological dynamics}\label{app3}
The complete list of equations for the population dynamics is

\begin{flalign}
\frac{dy_j^S}{dt}&=-p y_j^S\sum_{i'}x_{i'}^F +\sum_{i'}q_j P_{i'j}+(1-\theta)\sum_{i'}s_{i'}P_{i'j}+\frac{1}{h}y_j^H+\gamma y_j^H-dy_j^S,\\
\frac{dy_j^H}{dt}&=\nu p y_j^S\sum_{i'}x_{i'}^F+\theta \sum_{i'} s_{i'}P_{i'j} -\frac{1}{h}y_j^H-dy_j^H,\\
\frac{dx_i^F}{dt}&=r x_i^F \left(1-\frac{\sum_{i'} x_{i'}^F}{K}\right)-px_i^F\sum_{j'}y_{j'}^S+\sum_{j'}q_{j'}P_{ij'}+(1-\theta)\sum_{j'} s_i P_{ij'},\\
\frac{dP_{ij}}{dt}&=(1-\nu)px_i^Fy_j^S-(q_j+s_i)P_{ij},\\
\frac{dx_i}{dt}&=r x_i^F \left(1-\frac{\sum_{i'} x_{i'}^F}{K}\right)-\nu p x_{i}^F\sum_{j'}y_{j'}^S-\theta \sum_{j'} s_iP_{ij'},\\
\frac{dy_j}{dt}&=\gamma y_j^H - dy_j
\end{flalign}
with the conservation laws in (\ref{conslawpred}) and (\ref{conslawprey}). \\

We separate the dynamics into two time-scales by introducing the small and dimensionless scaling parameter $\varepsilon >0$: $y_j^S=\varepsilon \tilde{y}_{j}^S$, $y_j^H=\varepsilon \tilde{y}_{j}^H$, $y_j=\varepsilon \tilde{y}_{j}$, $P_{ij}=\varepsilon \tilde{P}_{ij}$, $p=\varepsilon^{-1} \tilde{p}$, $q_j=\varepsilon^{-1} \tilde{q}_{j}$, $s_i=\varepsilon^{-1} \tilde{s}_{i}$, $h=\varepsilon \tilde{h}$. We give the slow-fast equations using the scaled variables and parameters:

\begin{flalign}
\frac{d\tilde{y}_j^S}{dt}&=-\varepsilon^{-1}\tilde{p} \tilde{y}_j^S\sum_{i'}x_{i'}^F +\sum_{i'}\varepsilon^{-1} \tilde{q}_{j}\tilde{P}_{i'j}+(1-\theta)\sum_{i'}\varepsilon^{-1} \tilde{s}_{i'}\tilde{P}_{i'j}+\varepsilon^{-1}\frac{1}{ \tilde{h}}\tilde{y}_j^H+\gamma \tilde{y}_j^H-d\tilde{y}_j^S,\\
\frac{d\tilde{y}_j^H}{dt}&=\nu \varepsilon^{-1} \tilde{p} \tilde{y}_j^S\sum_{i'}x_{i'}^F+\theta \sum_{i'} \varepsilon^{-1} \tilde{s}_{i'}\tilde{P}_{i'j} -\varepsilon^{-1}\frac{1}{ \tilde{h}}\tilde{y}_j^H-d\tilde{y}_j^H,\\
\frac{dx_i^F}{dt}&=r x_i^F \left(1-\frac{\sum_{i'} x_{i'}^F}{K}\right)- \tilde{p} x_i^F\sum_{j'}\tilde{y}_{j'}^S+\sum_{j'} \tilde{q}_{j'}\tilde{P}_{ij'}+(1-\theta)\sum_{j'} \tilde{s}_{i} \tilde{P}_{ij'},\\
\frac{d\tilde{P}_{ij}}{dt}&=(1-\nu) \varepsilon^{-1} \tilde{p} x_i^F\tilde{y}_j^S-\varepsilon^{-1}( \tilde{q}_{j}+\tilde{s}_{i})\tilde{P}_{ij},\\
\frac{dx_i}{dt}&=r x_i^F \left(1-\frac{\sum_{i'} x_{i'}^F}{K}\right)-\nu \tilde{p} x_{i}^F\sum_{j'}\tilde{y}_{j'}^S-\theta \sum_{j'}\tilde{s}_i \tilde{P}_{ij'},\\
\frac{d \tilde{y}_j}{dt}&=\gamma  \tilde{y}_j^H - d \tilde{y}_j.
\end{flalign}

We introduce the short time $\tilde{t}=\varepsilon^{-1} t$, let $\varepsilon \rightarrow 0$ and drop the tildes to obtain the equations for the fast dynamics

\begin{flalign}
\frac{dy_j^S}{dt}&=-p y_j^S\sum_{i'}x_{i'}^F +\sum_{i'}q_j P_{i'j}+(1-\theta)\sum_{i'}s_{i'}P_{i'j}+\frac{1}{h}y_j^H,\\
\frac{dy_j^H}{dt}&=\nu p y_j^S\sum_{i'}x_{i'}^F+\theta \sum_{i'} s_{i'}P_{i'j} -\frac{1}{h}y_j^H,\\
\frac{dx_i^F}{dt}&=0,\\
\frac{dP_{ij}}{dt}&=(1-\nu)px_i^Fy_j^S-(q_j+s_i)P_{ij},\\
\frac{dx_i}{dt}&=0,\\
\frac{dy_j}{dt}&=0.
\end{flalign}

We use the conservation law for the total predator density to reduce the equations for $y_j^S$, $y_j^H$ and $P_{ij}$ to only two equations and we set $x_i^F=x_i$. We obtain the equations for the quasi-equilibrium from (\ref{predequno}) and (\ref{paireq}):

\begin{flalign}\label{fast1}
0&=-p x y_j^S +\sum_{i'}q_j P_{i'j}+(1-\theta)\sum_{i'}s_{i'}P_{i'j}+\frac{1}{h}(y_j-y_j^S-\sum_{i'}P_{i'j}),\\\label{fast2}
0&=(1-\nu)px_iy_j^S-(q_j+s_i)P_{ij}.
\end{flalign}

We conclude that the system has a unique quasi-equilibrium of the fast dynamics,

\begin{flalign}\label{quasiF}
x_i^F&=x_i, \\ \label{quasiS}
y_j^S&=\frac{y_j}{1+hpx+p (\nu-1) \left[(hq_j-1) \sum_{i'} \frac{x_{i'}}{q_j+s_{i'}} + h(1-\theta) \sum_{i'} \frac{s_{i'}x_{i'}}{q_j+s_{i'}} \right]},\\ \label{quasiP}
P_{ij}&=\frac{(1-\nu)\frac{p}{q_j+s_i}x_iy_j}{1+hpx+p (\nu-1) \left[(hq_j-1) \sum_{i'} \frac{x_{i'}}{q_j+s_{i'}} + h(1-\theta) \sum_{i'} \frac{s_{i'}x_{i'}}{q_j+s_{i'}} \right]},\\ \label{quasiH}
y_j^H&=y_j-y_j^S-\sum_{i'}P_{i'j}.
\end{flalign}

When the denominator of (\ref{quasiS}) and (\ref{quasiP}) is positive, then the quasi-equilibrium is feasible, i.e. exists and is positive. Furthermore we use linear stability analysis to check the stability conditions. The elements of the Jacobian matrix corresponding to the fast system of equations in (\ref{fast1}) and (\ref{fast2}) are constant and do not depend on the equilibrium, thus computing the trace and determinant and their signs is straightforward. We conclude that under the feasibility condition, the trace and determinant are respectively negative and positive, hence the quasi-equilibrium is hyperbolically stable. To prove global stability we use the Poincar\'e-Bendixon theorem and the Bendixon-Dulac theorem. We can exclude the existence of a limit cycle, because the trace is negative everywhere. Furthermore, since there is only one equilibrium, it must be the $\omega$-limit of every orbit of the fast dynamics.

\section{On the equilibrium for the ecological dynamics}\label{app2}
We consider the differential equations in (\ref{ecodynx}) and (\ref{ecodyny}) for a single prey $x$ and a single predator $y$ with strategies respectively $s$ and $q$ and $\beta_{s,q}$ and $h_{s,q}$ as defined for (\ref{fronetype}),
\begin{eqnarray}
\dot{x}&=&x (1-x) -\frac{\beta_{s,q} x}{1+\beta_{s,q}h_{s,q}x},\\
\dot{y}&=&\gamma \frac{\beta_{s,q}xy}{1+\beta_{s,q} h_{s,q}x}-dy.
\end{eqnarray}

The prey zero-growth and predator zero-growth isoclines (see Figure \ref{fig:sec1}) are

\begin{flalign}
x=&0,\\ \nonumber
y=&\frac{(1-x)(1+h_{s,q}\beta_{s,q}x)}{\beta_{s,q}}=\\
=&\frac{(x-1)[q-px(\nu-1)+hpq\nu x+s(1+hp\theta x + p\nu x-p\theta\nu x)]}{p[s\theta(\nu-1)-q\nu-s\nu]},\\
y=&0,\\\nonumber
x=&\frac{d}{\beta_{s,q}(\gamma-d h_{s,q})}=\\
=&\frac{d (q+s)}{pd [-hq\nu-hs (\theta +\nu-\nu\theta)+\nu -1]+p\gamma  [\nu  q+s (\theta +\nu-\nu\theta)]}.\end{flalign}

The full expressions for interior equilibrium in (\ref{ecoeq}) is 

\begin{flalign}\label{fullx}
\hat{x}=&\frac{d (q+s)}{pd [-hq\nu-hs (\theta +\nu-\nu\theta)+\nu -1]+p\gamma  [\nu  q+s (\theta +\nu-\nu\theta)]},\\\label{fully}
\hat{y}=&\frac{(q+s) p\gamma^2[q\nu+s(\theta+\nu-\theta\nu)]-(q+s)d \gamma[q+s+p-p\nu+pq\nu+ps(\theta+\nu-\theta\nu)]}{p^2[q\nu\gamma+s\gamma(\theta+\nu-\theta\nu)-d+d\nu-hdq\nu-hds(\theta+\nu-\theta\nu)]^2}.
\end{flalign}

The equilibrium is viable if $\gamma>dh$ and one of the following sets of conditions for $p$, $s$ and $q$ holds

\begin{flalign}\nonumber
&\frac{d}{(\gamma-dh)(\theta+\nu-\theta\nu) }<p<\frac{d}{(\gamma-dh)\nu},\\\nonumber
&s>\frac{dp(1-\nu)}{d h \theta  \nu  p-hd \theta  p-d h \nu  p-d-\gamma  \theta  \nu  p+\gamma  \theta  p+\gamma  \nu  p},\\\label{viab1}
&0<q<\frac{d p [\nu -1-h s ( \theta+\nu-\theta\nu )]-d s+\gamma  p s (\theta +\nu- \theta\nu)}{d h \nu  p+d-\gamma  \nu  p},
\end{flalign}
or

\begin{flalign}\label{viab2}
p= \frac{d}{(\gamma-dh)\nu}, \quad s>\frac{dp(1-\nu)}{d h \theta  \nu  p-hd \theta  p-d h \nu  p-d-\gamma  \theta  \nu  p+\gamma  \theta  p+\gamma  \nu  p}, \quad q>0,
\end{flalign}
or

\begin{flalign}\nonumber
&p> \frac{d}{(\gamma-dh)\nu}, 
\quad 0<s<\frac{dp(1-\nu)}{-d-dhp\theta+p\gamma\theta-dhp\nu+p\gamma\nu+dp\theta\nu-p\gamma\theta\nu}, \\\label{viab3}
&q>\frac{d p [\nu -1-h s ( \theta+\nu-\theta\nu )]-d s+\gamma  p s (\theta +\nu- \theta\nu)}{d h \nu  p+d-\gamma  \nu  p},
\end{flalign}
or

\begin{flalign}\label{viab4}
p> \frac{d}{(\gamma-dh)\nu}, \quad
s\geq\frac{dp(1-\nu)}{-d-dhp\theta+p\gamma\theta-dhp\nu+p\gamma\nu+dp\theta\nu-p\gamma\theta\nu}, 
\quad q>0.
\end{flalign}
Therefore in the positive quadrant of the $(s,q)$-plane, the interior equilibrium exists and is positive for values of $(s,q)$ on the right-hand side of the \emph{extinction boundary} $q=\frac{d p [\nu -1-h s ( \theta+\nu-\theta\nu )]-d s+\gamma  p s (\theta +\nu- \theta\nu)}{d h \nu  p+d-\gamma  \nu  p}$, which intersects the $s$-axis at $s=\frac{d (1-\nu) p}{p(\gamma -dh) ( \theta+\nu-\theta\nu )-d}$ (see Figure \ref{fig:sec21} and Figure \ref{fig:sec22}) for different configurations). A transcritical bifurcation occurs on the extinction boundary: for values of $(s,q)$ on the left-hand side of the line, the interior equilibrium is no longer present and the dynamics converges to the predator-free equilibrium.\\

The determinant and the trace of the community matrix $J(x,y)\Big|_{x=\hat{x},y=\hat{y}}$ for the ecological dynamics determine the stability of the equilibrium:

\begin{flalign}
\det J(x,y)\Big|_{x=\hat{x},y=\hat{y}} &= d(2 \hat{x}-1) -\frac{\beta_{s,q}[\hat{x}(2 \hat{x}-1)-d \hat{y} +\gamma h_{s,q} \beta_{s,q} \hat{x}^2 (2 \hat{x}-1)]}{(1+h_{s,q} \beta_{s,q} \hat{x})^2}\\
\text{tr} J(x,y)\Big|_{x=\hat{x},y=\hat{y}} &=1-d-2 \hat{x}+\frac{\beta_{s,q}(\gamma\hat{x}-\hat{y}+\gamma h_{s,q} \beta_{s,q} \hat{x}^2)}{(1+h_{s,q} \beta_{s,q} \hat{x})^2}
\end{flalign}
A Hopf bifurcation occurs when the trace evaluated at the interior equilibrium is zero. In order to verify that the Hopf bifurcation do not coincide with the transcritical bifurcation, we look at the $(s,q)$-planes in Figure \ref{fig:sec21} and Figure \ref{fig:sec22}: it is enough to check that the two curves cannot intersect with the $s$-axis in the same point. By solving for different parameters the equation for the intersection points of, respectively, the Hopf bifurcation line and the extinction border with the axis $q=0$, we obtain that under no conditions on the parameter values the two bifurcations coincide.

\section{On the sign of the predator gradient}\label{app4}
In this Section we discuss the proof of Proposition \ref{prop1}. 
The condition $\hat{x}=\frac{s\theta}{p\nu}$ in (\ref{gradcond2}) is equivalent to $(\nu p \hat{x})^{-1}=(\theta s)^{-1}$ and compares the rate of the two events modelling successful prey capture by the predator (here given in the form of monomolecular reactions with a constant or prey density-dependent transition rate) with $x^F=x$ at the fast time equilibrium

\begin{flalign}\nonumber
& \mathcircled{y^S}   \xrightarrow{\nu p x} \mathcircled{y^H}  \quad   \textit{attack and prey capture}, \\ \nonumber
& \mathcircled{P}  \xrightarrow{\theta s}  \mathcircled{y^H} \quad  \textit{prey ends stand-off and is captured}. \\ \nonumber
\end{flalign}
In particular, let's define with $ES$ the expected time till prey capture for a predator starting in state $y^S$ and with $EP$ the expected time till prey capture for a predator starting in state $P$. Then, following the definitions, we obtain the system of equations for $ES$ and $EP$
\begin{eqnarray}
ES&=&\frac{1}{px}+\frac{(1-\nu)px}{\nu px+ (1-\nu)px}EP,\\
EP&=&\frac{1}{q+s}+\frac{q+(1-\theta)s}{q+s}ES.
\end{eqnarray}
Solutions for $ES$ and $EP$ are
\begin{eqnarray}
ES&=&\frac{q+s+px(1-\nu)}{px[q\nu+s(\theta+\nu-\theta\nu)]},\\
EP&=&\frac{q+s+px-\theta s}{px[q\nu+s(\theta+\nu-\theta\nu)]},
\end{eqnarray}
the comparison of the two ($ES=EP$) leads to condition (\ref{gradcond2}). As $\hat{x}=\frac{s\theta}{p\nu}$ can be reformulated in terms of $s^*$, we conclude that if $s=s^*$, then $ES=EP$. In the same way, starting from the conditions in (\ref{gradcond1}) and (\ref{gradcond3}), we get that if $s>s^*$, then $ES>EP$, and if $s<s^*$, then $ES<EP$.\\

\section{On the canonical equation and different definitions of stability}\label{app1}
We recall the canonical equations in (\ref{evodyns}) and (\ref{evodynq}). 
For the purpose of this study, below we only give conditions for stability of a singularity in the monomorphic resident population and we assume that the coefficients $k_{i}(s,q), i=prey,pred$ are differentiable. We can define an evolutionary attractor by analysing the trace and determinant of the Jacobian matrix of the evolutionary dynamics in (\ref{evodyns}) and (\ref{evodynq}) evaluated at the singularity. The elements of $J(s,q)\Big|_{s=s^*,q=q^*}$ are

\begin{flalign}
J_{11}&=k_{prey}(s,q)\left( \frac{\partial^2 g_{prey}(s_m,q)}{\partial s_m^2}+\frac{\partial^2g_{prey}(s_m,q)}{\partial s\partial s_m} \right)=k_{prey}(s,q) \mathcal{A}_{11},\\
J_{12}&=k_{prey}(s,q) \frac{\partial^2 g_{prey}(s_m,q)}{\partial s_m \partial q}=k_{prey}(s,q) \mathcal{A}_{12},\\
J_{21}&=k_{pred}(s,q) \frac{\partial^2 g_{pred}(s,q_m)}{\partial q_m \partial s}=k_{pred}(s,q) \mathcal{A}_{21},\\
J_{22}&=k_{pred}(s,q) \left( \frac{\partial^2g_{pred}(s,q_m)}{\partial q_m^2}+\frac{\partial^2g_{pred}(s,q_m)}{\partial q\partial q_m} \right)=k_{pred}(s,q) \mathcal{A}_{22}.
\end{flalign}
By the Routh-Hurwitz criterion, a singularity (more generally, a singular coalition of one-dimensional traits) is \emph{convergence stable} if $\det(J)>0$ and $\operatorname{tr}(J)<0$. A singularity $(s^*,q^*)$ is \emph{weakly convergence stable} if there exist a given strictly positive diagonal matrix with diagonal entries $k_{i}(s,q)>0, i=prey,pred$ such that

\begin{flalign}\label{weak1}
&k_{prey}(s,q)\mathcal{A}_{11}+k_{pred}(s,q) \mathcal{A}_{22}<0,\\\label{weak2}
&\mathcal{A}_{11}\mathcal{A}_{22}>\mathcal{A}_{12}\mathcal{A}_{21}.
\end{flalign}
For the present model, we have \emph{weak stability} if and only if

\begin{flalign}\label{weak1here}
&\mathcal{A}_{11}<0 \lor \mathcal{A}_{22}<0,\\\label{weak2here}
&\mathcal{A}_{11}\mathcal{A}_{22}>\mathcal{A}_{12}\mathcal{A}_{21},
\end{flalign}
as we suppose either $k_{prey}(s,q)$ or $k_{pred}(s,q)$ large enough to make the other term in (\ref{weak1}) negligible.
\emph{Strong convergence stability} is obtained for any strictly positive diagonal matrix with diagonal entries $k_{i}(s,q)>0, i=prey,pred$ and

\begin{flalign}\label{strong1}
&\mathcal{A}_{11}<0, \quad \mathcal{A}_{22}<0,\\\label{strong2}
&\mathcal{A}_{11}\mathcal{A}_{22}>\mathcal{A}_{12}\mathcal{A}_{21}.
\end{flalign}
Note that strong convergence stability implies weakly convergence stability. Furthermore, we define a strategy $(s^*,q^*)$ \emph{totally stable} if it is stable for the differential inclusions

\begin{eqnarray}
0\leq\frac{d}{dt}(s-s^*) (\mathcal{A}_{11}(s-s^*)+\mathcal{A}_{12}(q-q^*)),\\
0\leq\frac{d}{dt}(q-q^*) (\mathcal{A}_{21}(s-s^*)+\mathcal{A}_{22}(q-q^*)).
\end{eqnarray}
In particular, $(s^*,q^*)$ is totally stable if

\begin{flalign}\label{totally1}
&\mathcal{A}_{11}<0, \quad \mathcal{A}_{22}<0,\\\label{totally2}
&\mathcal{A}_{11}\mathcal{A}_{22}>\left|\mathcal{A}_{12}\mathcal{A}_{21}\right|
\end{flalign}
and total stability implies strong stability of the singular strategy (see also \cite{matessi1996long}).

\section*{Acknowledgements}
This research was funded by the Academy of Finland, Centre of Excellence in Analysis and Dynamics Research.

\bibliographystyle{abbrv}  
\bibliography{myref}
\nocite{*}

\end{document}